\documentclass[prl,reprint,aps,tightenlines,preprintnumbers,amsmath,amssymb,
showpacs,superscriptaddress]{revtex4-1}
\usepackage{amsmath,amssymb,graphicx,natbib}
\usepackage{graphicx}
\usepackage{dcolumn}
\usepackage{bm}
\usepackage[hyperindex,breaklinks]{hyperref}
\usepackage{amsmath}
\usepackage{array}

\def\be{\begin{equation}}
\def\ee{\end{equation}}
\def\ba{\begin{eqnarray}}
\def\ea{\end{eqnarray}}
\def\ge{\mathrel{\raise.3ex\hbox{$>$\kern-.75em\lower1ex\hbox{$\sim$}}}}
\def\la{\mathrel{\raise.3ex\hbox{$<$\kern-.75em\lower1ex\hbox{$\sim$}}}}

\def\simgt{\mathrel{\raise.3ex\hbox{$>$\kern-.75em\lower1ex\hbox{$\sim$}}}}
\def\simlt{\mathrel{\raise.3ex\hbox{$<$\kern-.75em\lower1ex\hbox{$\sim$}}}}

\newcommand{\fr}[2]{\frac{#1}{#2}}

\newcommand{\nc}{\newcommand}

\nc{\gone}{\bar g_{\pi NN}^{(1)}}
\nc{\gzero}{\bar g_{\pi NN}^{(0)}}
\nc{\al}{\alpha}
\nc{\ga}{\gamma}
\nc{\de}{\delta}
\nc{\ep}{\epsilon}
\nc{\ze}{\zeta}
\nc{\et}{\eta}

\nc{\Th}{\Theta}
\nc{\ka}{\kappa}
\nc{\rh}{\rho}
\nc{\si}{\sigma}
\nc{\ta}{\tau}
\nc{\up}{\upsilon}
\nc{\ph}{\phi}
\nc{\ch}{\chi}
\nc{\ps}{\psi}
\nc{\om}{\omega}
\nc{\Ga}{\Gamma}
\nc{\De}{\Delta}
\nc{\La}{\Lambda}
\nc{\Si}{\Sigma}
\nc{\Up}{\Upsilon}
\nc{\Ph}{\Phi}
\nc{\Ps}{\Psi}
\nc{\Om}{\Omega}
\nc{\ptl}{\partial}
\nc{\del}{\nabla}
\nc{\ov}{\overline}
\nc{\newcaption}[1]{\centerline{\parbox{15cm}{\caption{#1}}}}

\def\beq{\begin{equation}}
\def\eeq{\end{equation}}
\def\bmat{\begin{displaymath}}
\def\emat{\end{displaymath}}
\def\bear{\begin{eqnarray}}
\def\eear{\end{eqnarray}}
\def\bery{\begin{array}}
\def\ery{\end{array}}
\def\bit{\begin{itemize}}
\def\eit{\end{itemize}}
\def\ben{\begin{enumerate}}
\def\een{\end{enumerate}}
\def\btab{\begin{tabular}}
\def\etab{\end{tabular}}
\def\btbl{\begin{table}}
\def\etbl{\end{table}}
\def\bfig{\begin{figure}[htb]}
\def\efig{\end{figure}}
\def\bpic{\begin{picture}}
\def\epic{\end{picture}}

\def\ga{\mathrel{\raise.3ex\hbox{$>$\kern-.75em\lower1ex\hbox{$\sim$}}}}
\def\la{\mathrel{\raise.3ex\hbox{$<$\kern-.75em\lower1ex\hbox{$\sim$}}}}
\def\gappeq{\mathrel{\rlap {\raise.5ex\hbox{$>$}}
{\lower.5ex\hbox{$\sim$}}}}
\def\lappeq{\mathrel{\rlap{\raise.5ex\hbox{$<$}}
{\lower.5ex\hbox{$\sim$}}}}

\def\gyr{{\rm \, G\kern-0.125em yr}}
\def\mev{{\rm \, Me\kern-0.125em V}}
\def\gev{{\rm \, Ge\kern-0.125em V}}
\def\tev{{\rm \, Te\kern-0.125em V}}

\begin{document}

\title{Testing Parity with Atomic Radiative Capture of $\mu^-$}

\author{David McKeen}
\affiliation{Department of Physics and Astronomy, University of Victoria, 
Victoria, BC V8P 5C2, Canada}

\author{Maxim Pospelov}
\affiliation{Department of Physics and Astronomy, University of Victoria, 
Victoria, BC V8P 5C2, Canada}
\affiliation{Perimeter Institute for Theoretical Physics, Waterloo, ON N2J 2W9, 
Canada}

\begin{abstract}

The next generation of ``intensity frontier" facilities will bring a significant
increase in the intensity of sub-relativistic beams of $\mu^-$. We show that the
use of these beams in combination with thin targets of $Z\sim 30$ elements opens
up the possibility of testing parity-violating interactions of muons with nuclei
via direct radiative capture of muons into atomic $2S$ orbitals. Since atomic 
capture preserves longitudinal muon polarization, the measurement of the gamma 
ray angular asymmetry in the single photon $2S_{1/2}$--$1S_{1/2}$ transition 
will offer a direct test of parity. We calculate the probability of atomic 
radiative capture taking into account the finite size of the nucleus to show 
that this process can dominate over the usual muonic atom cascade, and that the 
as yet unobserved single photon $2S_{1/2}$--$1S_{1/2}$ transition in muonic 
atoms can be detected in this way using current muon facilities.

\end{abstract}

\maketitle

\newpage

{\em Introduction.}---The standard model of particles and fields (SM) has shown 
tremendous vitality under an onslaught of new TeV-scale data from the Large 
Hadron Collider (LHC). Stringent limits are derived on new hypothetical vector 
particles $Z'$ that mediate interactions between light quarks and charged 
leptons. For a sequential SM $Z$-like $Z'$ particle such limits extend to 2 TeV,
rendering low-energy parity-violating tests not competitive with the LHC in the 
search for new heavy resonances with large couplings to SM particles. However, 
an alternative possibility--light and very weakly coupled particles--may easily 
escape the high-energy constraints while inducing some nontrivial effects at low
energy~\cite{Fayet,*Fayet2}. In recent years the interest in this type of 
physics has intensified, largely due to the accumulation of various anomalous 
observations that such light particles may help to explain. (For a possible 
connection between light vectors and dark matter physics see, {\em e.g.}, 
Ref.~\cite{lightV,*lightV2}.) In parallel with this, attempts to detect such new 
states at ``intensity frontier" facilities are becoming more frequent and more 
systematic~\cite{A1,*Apex}. 

Muon physics, and its study with new high intensity muon beams, is a natural 
point of interest because of the lingering discrepancy between calculations and 
measurements of the muon anomalous magnetic moment~\cite{g-2} as well as the 
recent striking discrepancy of the proton charge radius extracted from the 
muonic hydrogen Lamb shift~\cite{muH} as compared to other determinations of the
same quantity~\cite{CODATA,*HillPaz}. While it is far from clear that these 
discrepancies are not caused by some poorly understood SM physics or 
experimental mistakes, it is still important to investigate models of New 
Physics (NP) that could create such deviations. Models with light vector 
particles (see, {\em e.g.},~\cite{Okun,*Holdom}) are particularly interesting as
they can remove the $g-2$ discrepancy quite naturally~\cite{Gninenko,*Pospelov},
or be responsible for extra muon-proton interactions that can be interpreted as 
a shift of the proton charge radius~\cite{newforce1,*newforce2,*newforce3,BMP}.

As was argued in Ref.~\cite{BMP}, a lepton flavor-specific muon-proton 
interaction in combination with constraints in the neutrino sector may imply 
that right-handed muon number is gauged, leading to new parity-violating 
muon-proton neutral current interactions. We take this model as a representative
example of new physics at the sub-GeV energy scale that can create stronger-than-weak 
effects in the interaction of muons with nuclei. In this Letter, we 
revisit the idea of searching for parity violation in the muon sector using 
muonic atoms, keeping in mind that no direct tests of the axial vector muon 
coupling have been performed at low energy, and that the NP contribution could 
dominate over the SM~\cite{BMP}. To be specific, we consider a low-energy 
effective neutral current Lagrangian, that includes the sum of the SM and NP 
contributions,
\begin{align}
\nonumber
&{\cal L}_\mu = {\cal L}_{\rm SM} + {\cal L}_{\rm NP}\\
\label{SM} 
&{\cal L}_{\rm SM}=-\fr{G_F}{2\sqrt{2}}
\bar \mu\gamma_\nu\gamma_5 \mu \left(g_n\bar n \gamma_\nu n + g_p
\bar p \gamma_\nu p\right),\\
\label{NP} 
&{\cal L}_{\rm NP}=
\bar \mu\gamma_\nu\gamma_5 \mu \fr{4\pi\alpha g^{\rm NP}_\mu}{m_V^2+\Box}
\left(g^{\rm NP}_n\bar n \gamma_\nu n + g^{\rm NP}_p
\bar p \gamma_\nu p\right)
\end{align}
where the SM vector couplings to nucleons are given by $g^V_n = -\fr12 $, 
$g^V_p = \fr12-2\sin^2\theta_W $. In the model with gauged right-handed muon 
number, the least constrained points in the parameter space correspond to the 
mass of the mediator gauge boson of $m_V \simeq 30$ MeV. In that case, the fit 
to the proton charge radius suggests~\cite{BMP}
\be
\fr{4\pi\alpha g_\mu^{\rm NP}g^{\rm NP}_p}{m_V^2} \simeq 
\frac{2\times 10^{-5}}{(30~{\rm MeV})^2}\gg G_F,
\ee
which should be considered as perhaps the most optimistic value for the strength
of the muon-proton interaction. In what follows we suggest a new way to search 
for the manifestation of (\ref{SM}) and (\ref{NP}) in muonic atoms using the 
process of atomic radiative capture (ARC) to the $2S$ state: 
$\mu^- +Z \to (\mu^- Z)_{2S} +\gamma$. We show that probing ${\cal L}_{\rm NP}$ 
of maximal strength is possible with existing muon line facilities, while the SM
values can eventually be tested at the next generation of high-intensity muon 
sources. 

\begin{figure}
\rotatebox{0}{\resizebox{40mm}{!}{\includegraphics{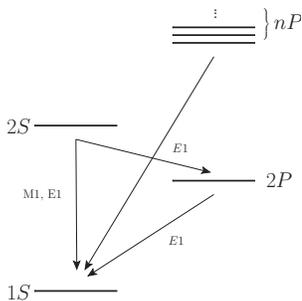}}}
\caption{A diagram of the atomic levels in typical muonic atoms.  Also shown are
some of the single photon transitions between states. The $2S\to1S$ single 
photon transition is an admixture of a suppressed M1 transition and an E1
transition from $2S$-$2P$ mixing induced by parity violation.}
\label{fig:levels}
\end{figure}

It is well-known that the suppressed M1 single photon $2S_{1/2}$--$1S_{1/2}$ 
transition in combination with the small energy difference between the $2S$ and 
$2P$ states enhances the parity-violating asymmetry in M1-E1 interference.  This
idea has received a significant amount of theoretical and experimental 
attention, summarized in the review~\cite{MS}. The most promising scheme for the
detection of parity violation to date was identified as a slow muon forming a 
highly excited atomic state with a nucleus followed by a cascade ending with
\begin{eqnarray}
...\to2S_{1/2}\xrightarrow{M1-E1}1S_{1/2}+\gamma;~(\mu^-)_{1S} 
\to e^- \nu_\mu\bar\nu_e,
\label{oldscheme}
\end{eqnarray}
with parity violation being encoded in the correlation between the directions of
the outgoing $\gamma$ and the muon decay electron. In Fig.~\ref{fig:levels} we 
show a level diagram for a typical muonic atom.

Despite considerable efforts, the single photon $2S$--$1S$ transition itself has
never been detected in any muonic atoms. In light atoms, $Z\la 10$, this 
transition cannot be distinguished from the far more dominant $2P$--$1S$, as the
difference between gamma ray energies in this case is much smaller than the 
energy resolution of $\gamma$-detectors. Combining this with the tiny branching 
ratio of the one-photon decay of the $2S_{1/2}$ state in light elements, and the
fact that it gets scarcely populated, $O(1\%)$, during the cascade, makes the 
measurement of parity violation very challenging in light muonic atoms, even 
though the value of parity-violating asymmetries could be as large as few 
percent~\cite{MS}.
Heavier muonic atoms, $Z\sim 30$, have been suggested as promising candidates to
test parity~\cite{CF}, because the $2S$--$1S$ and $2P$--$1S$ transitions can be 
easily resolved, as the energy difference between the $2S$ and $2P$ states 
reaches
\begin{align}
\label{DeltaE}
\Delta E &\equiv  E_{2S}-E_{2P} = \frac{(Z\alpha)^4m_\mu(m_\mu R_c)^2}{12}\\
\nonumber
&\simeq 210~{\rm keV} \times (Z/36)^4\times (R_c/4.2~{\rm fm})^2,
\end{align}
where we have normalized the nuclear charge, $Z$, and the nuclear charge radius,
$R_c$, on the values for krypton, and suppressed total $J$ indices, effectively 
neglecting the splitting between $2P_{3/2}$ and $2P_{1/2}$ states. 
Unfortunately, as in the case of lighter elements, the $2S$--$1S$ transition was
never detected in heavier atoms, because of the dominance of the background 
created by quanta from $nP$--$1S$ transitions, $n\geq 3$, whose energies have 
been degraded~\cite{MS}. To elaborate on this, one can estimate the signal-to-
background ratio of the single photon $2S$--$1S$ transition during the atomic 
cascade. The signal, $S\sim N_{2S}{\rm Br}_{1\gamma}$, is proportional to the 
fraction of cascade muons $N_{2S}$ that end up in the $2S$ state, where $N_{2S}$
is typically on the order of $10^{-2}$~\cite{Klaus}, and the branching of M1 
single photon transition from $2S$ states, which for $Z\sim 30 $~\cite{CF} is 
given by 
\begin{align}
\label{Br}
{\rm Br} _{1\gamma}&\simeq \frac{\Gamma_{2S-1S+1\gamma}}
{\Gamma_{2S-2P}+\Gamma_{2S-1S+2\gamma}+\Gamma_{\rm Auger}}
\\
&\simeq \frac{\Gamma_{2S-1S+1\gamma}}{\Gamma_{2S-2P}}  \sim  2\times 10^{-3}.
\nonumber
\end{align}
For smaller $Z$, $Z<28$, the single photon branching is strongly suppressed by 
Auger processes~\cite{GS} and by the two photon transitions. The cascade-related
background consists of the number of energy-degraded $nP$--$1S$ ($n\geq 3$) 
photons ({\em i.e.} those that do not deposit their full energy in the detector)
that fall into the energy resolution interval $\Delta E$ centered at the energy 
of the $2S$--$1S$ transition. From experimental studies~\cite{cascade1,
*cascade2} one can conclude that $O(20\%)$ of muons undergoing a cascade 
generate $nP$--$1S$ transitions. For realistic $\gamma$-detectors, the number of
energy-degraded photons is $\sim50\%$, and the number of photons under the 
$2S$--$1S$ peak within the energy resolution window of $\Delta E\sim 2$~keV 
can be estimated as  $ B \sim 0.2\times \Delta E/(2E_\gamma) \sim  10^{-4}$ for 
$E_\gamma\sim 2$~MeV. Therefore, one arrives at the following estimate of signal
-to-background:
\be
\left[ \fr SB \right]_{\rm cascade} \leq 0.2.
\label{SB_C}
\ee
The actual ratio is smaller than this upper bound because of additional photon 
backgrounds caused by other sources, which explains why the $2S$--$1S$ 
transition has not been detected~\cite{MS}.

In addition to these challenges in detecting the $2S$--$1S$ transition in muon 
cascades, another difficulty in implementing the scheme in (\ref{oldscheme}) 
lies in the fact that the final step, muon decay, for these elements is very 
subdominant to nuclear muon capture. Because of the combination of these two 
factors, parity experiments with $Z\sim 30$ elements were deemed impractical~
\cite{MS}. 

{\em New proposal for a parity-violation measurement.}--Our proposal is to 
abandon (\ref{oldscheme}) and use thin targets of $Z\geq 30$ elements that only 
decrease the $\mu^-$ momentum, but do not stop the particle completely. This 
removes most of the background related to the muonic cascade. A fraction of the 
muons undergo ARC directly into the $2S$ state.  The signal consists of two 
$\gamma$ quanta, one from the ARC process ($\gamma_1$), and the other from the 
single photon decay of the $2S$ state ($\gamma_2$):
\be
\mu^-_{\rightarrow} + Z \to (\mu^-_{\rightarrow} Z)_{2S_{1/2}} +\gamma_1;
~~2S_{1/2}\xrightarrow{M1-E1}1S_{1/2}+\gamma_2.
\label{newscheme}
\ee
Here $\mu^-_{\rightarrow}$ denotes the longitudinally polarized muon. While for 
the relevant range of $Z$ the energy of $\gamma_2$ is on the order of $2$ MeV, 
the energy of $\gamma_1$ is dependent on the muon momentum, and for muon 
momentum of $50 $ MeV is in the 10 MeV range. The parity-violating signature is 
the forward-backward asymmetry of $\gamma_2$ relative to the direction of the 
muon spin. 

\begin{figure}
\rotatebox{0}{\resizebox{50mm}{!}{\includegraphics{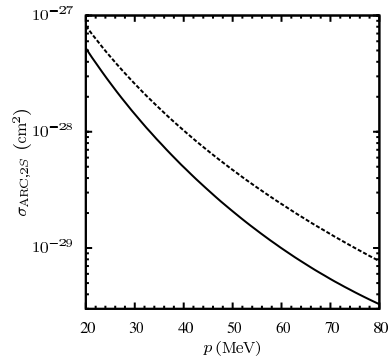}}}
\caption{$\sigma_{{\rm ARC},2S}$ as a function of the incoming muon momentum, 
$p$ (solid curve) for a muon scattering on krypton, $Z=36$, with a uniform 
nuclear charge density and charge radius of $R_c = 4.2$~fm while taking the 
departure from the dipole approximation into account. Also shown is the cross 
section in the dipole approximation with a point-like nucleus (dashed curve).  
ARC into the $2S$ state is a factor of a several less probable than into the 
$1S$ state.}
\label{fig:sigma}
\end{figure}

To calculate the cross section for muonic ARC into the $2S$ state (the first 
step in (\ref{newscheme})), we note that the analogous process involving an 
electron, electron-nucleus photorecombination, in the dipole approximation with 
a point-like nucleus is a standard textbook calculation~\cite{BS,LL}, as it can 
be obtained from the standard hydrogen-like photoelectric ionization cross 
section $\sigma_{PE}^{(0)}$. Here we adjust this for the muon case, which, 
besides the substitution $m_e\to m_\mu$, involves accounting for the finite 
nuclear charge radius and the departure from the dipole approximation. This can 
be done by introducing a correction factor to the standard formula,
\begin{align}
&\sigma_{\rm ARC} = \frac{2\omega^2}{p^2}\sigma_{PE};~\sigma_{PE} = 
\eta(p,R_c,Z,n,l) \times \sigma_{PE}^{(0)}(nl),
\nonumber
\\
& \sigma_{PE}^{(0)}(2S) = \frac{2^{14}\pi^2\alpha a^2E_2^4}{3\omega^4} 
\left[1+\frac{3E_2}{\omega}\right]\frac{\exp\{-\fr{4}{pa}\cot^{-1}
\fr{1}{2pa}\}}{1-\exp(-2\pi/pa)}.
\nonumber
\end{align}
In these expressions, $p$ is the momentum of the incoming muon, $a$ is the Bohr 
radius, $a= (Z\alpha m_\mu)^{-1}$, $E_2 = Z^2\alpha^2m_\mu/8$ is the 
(uncorrected) binding energy of the $2S$ muon, and $\omega = p^2/2m_\mu+E_2$ 
is the (uncorrected) energy of the photon emitted in the ARC process. The 
correction factor $\eta$ is calculated by  numerically solving the Schroedinger 
equation for a muon moving in the field of the nucleus with uniform charge 
distribution with charge radius $R_c$. The results for the cross sections are 
plotted in Fig.~\ref{fig:sigma} for $Z=36$ and $R_c = 4.2$~fm. As one can see, 
the corrections to the simple formula are significant, and mostly come from the 
finite charge of the nucleus, suppressing a naive cross section by more than a 
factor of $\sim 3$ for $p_\mu>60$ MeV. Moreover, at $p\sim m_\mu$, this formula 
will need to be further corrected by relativistic effects that thus far have 
been ignored in our treatment.

Previously, the ARC process was considered theoretically in Ref.~\cite{Chatt} 
for the case of muonic hydrogen, and searched for experimentally in Ref.~
\cite{cascade2} in muonic cascades in  Mg and Al. The ARC process was not 
detected because in the case of stopped muons the cross section for forming 
muonic atoms via electron ejection is several orders of magnitude larger than 
$\sigma_{\rm ARC}$. Because of that, one should not expect that the muon cascade
experiments can be sensitive to the ARC processes. 

Below, we estimate the probability for the ARC process in a thin gaseous target 
of Kr that decreases the momentum of the muon beam from 
$p_{\rm max}=30~{\rm MeV}$ to $p_{\rm min}=25~{\rm MeV}$:
\be
\label{parc}
P_{{\rm ARC},2S} = \int^{p_{\rm max}}_{p_{\rm min}} dp 
\frac{n_{\rm Kr}\sigma_{{\rm ARC},2S}}{|dp/dx|}\sim 2\times 10^{-7},
\ee
where the momentum loss, $dp/dx$, is given by standard Bethe-Bloch theory. For a
target size of $\sim 5$ cm, the number density of the krypton atoms would 
correspond to pressure of $p_{\rm Kr} \sim 8$~atm. 

Combining the probability of the ARC process (\ref{parc}) with the branching 
ratio of the M1 photons (\ref{Br}), we arrive at the emission rate of $2S$--$1S$
photons as a function of the incoming muon flux, 
\be
\label{rate}
\fr{d N_{2S-1S}}{dt} = P_{\rm ARC}\times {\rm Br}_{1\gamma}\times \Phi_{\mu^-} 
\sim\fr{1}{250~{\rm s}}\times  \fr{\Phi_{\mu^-}}{10^7 {\rm s}^{-1}}.
\ee
The lifetime of the $2S$ state is extremely small: for $Z>30$ it does not 
exceed $10~{\rm fs}$~\cite{CF} which allows for a tight timing correlation between $\gamma_1$ and $\gamma_2$ in~(\ref{newscheme}).

We can also estimate the intrinsic background created by the $nP$--$1S$ 
transitions in this case. For a transparent target, one source of background 
consists of the bremsstrahlung process, $\mu+Z\to \mu + Z +\gamma$ that degrades
the muon energy enough to trap it inside the target, with a subsequent muon 
cascade creating $nP$--$1S$ photons. To calculate the yield of $nP$--$1S$ 
photons, we estimate the probability for the process $\mu+Z\to \mu + Z +\gamma$
by taking the standard cross section \cite{LL} and modifying it by the 
correction coming from the finite nuclear charge. In this way we find, for the 
same parameters of the target, 
\be
P_{\rm cascade}\sim P_{\mu+Z\to \mu + Z +\gamma} \sim 20\times P_{{\rm ARC},2S},
\label{bkgnd}
\ee
requiring that the bremsstrahlung photon be at least as energetic as that coming from ARC into the $2S$ state for $p_{\rm min}=25~{\rm MeV}$.  Only a small fraction of the cascade photons, $\sim O(10^{-4})$, will be 
degraded to mimic the $2S$--$1S$ transition and we can conclude that the ratio 
of signal to irreducible background is 
\be
\left[\fr{S}{B}\right]_{\rm ARC} = \frac{P_{{\rm ARC},2S} \times 
{\rm Br}_{1\gamma}}{P_{\rm cascade}\times 10^{-4}} \sim O(1),
\label{SB_ARC}
\ee 
and the gain over (\ref{SB_C}) is rather significant. The contribution to the 
background due to direct capture on $n\geq 3$ orbits is even smaller. The background from bremsstrahlung and cascade photons in~(\ref{bkgnd}) is small enough that Ge detectors with $\mu{\rm s}$ response times can operate with muon fluxes of $O(10^{10}~{\rm s}^{-1})$ without photons from these processes arriving within the lifetime of the $2S$ state. We 
conclude that while the signal rate is small, Eq. (\ref{rate}), the gain in the 
$S/B$ can substantial, making the search for the ARC processes and $2S$--$1S$ 
transitions worth pursuing experimentally. A further increase in $S/B$ can be 
achieved by imposing a cut on the energy of $\gamma_1$ that can distinguish it 
from the lower energy bremsstrahlung $\gamma$. 

We are now ready to investigate the feasibility of the parity violation 
experiment with the use of the ARC scheme in (\ref{newscheme}). The forward-
backward asymmetry of the $2S$--$1S$ photon is related to the coefficient of 
$2S$--$2P$ mixing $\delta$ and the ratio of E1 and M1 amplitudes \cite{CF}, 
\begin{align}
\nonumber
&{\cal A}_{\rm FB}  = \fr{N_{\gamma_2}(\theta>\fr{\pi}{2})-N_{\gamma_2}(\theta<
\fr{\pi}{2})}
{N_{\gamma_2}(\theta>\fr{\pi}{2})+N_{\gamma_2}(\theta<\fr{\pi}{2})}=
2 \delta \fr{({\rm E1})_{2P-1S}}{({\rm M1})_{2S-1S}} \\
&\simeq 680 \times \left(  \frac{36}{Z} \right)^3\times 
\delta ,~
i\delta=\frac{\langle 2S_{1/2}|H_{PV} |2P_{1/2}\rangle}{\Delta E},
\end{align}
where the parity-violating Hamiltonian can be derived from (\ref{SM}) and 
(\ref{NP}). The size of the parity-violating admixture in the SM~\cite{CF} and 
in the presence of non-standard interactions~\cite{BMP}is given by
\begin{align}
&\delta_{\rm SM} \simeq \fr{3\sqrt{3}G_F }{8\sqrt{2}\pi Z \alpha R_c^2}
\left(g_p + g_n\frac{A-Z}{Z} \right),
\\
&\delta_{\rm NP} = \fr{3\sqrt{3}g_\mu^{\rm NP}}{2Z \alpha R_c^2m_\mu^2}
\fr{m_Va}{(m_Va+1)^3}\left(g_p^{\rm NP} + g_n^{\rm NP}\frac{A-Z}{Z} \right).
\nonumber
\end{align}
For the non-standard interaction (\ref{NP}), we normalize its strength to the 
possible size of the effect suggested by the muonic hydrogen Lamb shift 
discrepancy, following \cite{BMP}. This way for $Z=36$ we find 
\be
{\cal A}_{\rm FB}[{\rm SM}] \simeq 0.5\times 10^{-4},
~~~{\cal A}_{\rm FB}[{\rm NP}] = (0.5-11)\%.
\ee
The lower value of the asymmetry ${\cal A}_{\rm FB}[{\rm NP}]$ is for small, 
$\sim 10$ MeV, masses of vector mediators, while larger values are for the 
scaling regime, $m_V\gg 1/a$. 

Using these asymmetries and a realistic efficiency factor of $\sim 0.1$ for the 
detection of a two-photon transition, we arrive at the following estimate of the
time required to achieve the number of events 
$N\propto 1/{\cal A_{\rm FB}}^{2}$:
\begin{align}\nonumber
&T[{\rm SM}] \sim 10^8~{\rm s}\times\frac{10^{11}~{\rm s}^{-1}}{\Phi_\mu},
\\
&T[{\rm NP}] \sim 3\times 10^5~{\rm s}\times 
\frac{10^{7}~{\rm s}^{-1}}{\Phi_\mu}\times 
\left( \fr{0.1}{\cal A} \right)^2.
\end{align}
One can see that, while the test of a muonic parity violating 
${\cal A_{\rm FB}}$ down to the $O(10^{-4})$ value of the SM via the method 
suggested in this paper is statistically possible only with future high-
intensity muon beams, tests of some NP models \cite{BMP} are feasible even at 
existing facilities. 

In conclusion, let us summarize the main advantages of possible tests of parity 
using the atomic radiative capture scheme in Eq. (\ref{newscheme}):
\vspace{5pt}

\noindent {\em i.} The muon capture onto the $2S$ orbit proceeds via an E1 
transition and does not depolarize the muons. Therefore it is possible to 
capture a fully polarized muon onto the $2S$ orbit and study an angular 
asymmetry of the outgoing $\gamma$ without the need to observe muon beta decay 
in the $1S$ state. \\
{\em ii.} The gain in $S/B$ is significant, as the $nP$--$1S$ ($n>3$) 
transitions of cascade muons that prevented the detection of the single photon 
$2S$--$1S$ decay in the past are greatly reduced. The detection of this 
transition can be realistically performed even with the existing sources of 
$\mu^-$. \\
{\em iii.} The use of a transparent target allows one to study parity with muons
in a ``parasitic" set-up, when the dominant part of the muon flux is used for 
other experiments. It also appears that the ARC-based method (\ref{newscheme}) 
can withstand the increase of the muon beam intensity more easily than the 
cascade-based methods (\ref{oldscheme}). 

\vspace{5pt}

{\em Acknowledgments.}--We would like to thank A.~Antognini,  K.~Kirch, 
L.~Simons, C.~Svensson, and I.~Yavin for helpful discussions and communications. 
Research at the 
Perimeter Institute is supported in part by the Government of Canada through 
NSERC and by the Province of Ontario through MEDT. 

\vspace{-0.5cm}
\bibliography{muKr}

\begin{thebibliography}{27}%
\makeatletter
\providecommand \@ifxundefined [1]{%
 \@ifx{#1\undefined}
}%
\providecommand \@ifnum [1]{%
 \ifnum #1\expandafter \@firstoftwo
 \else \expandafter \@secondoftwo
 \fi
}%
\providecommand \@ifx [1]{%
 \ifx #1\expandafter \@firstoftwo
 \else \expandafter \@secondoftwo
 \fi
}%
\providecommand \natexlab [1]{#1}%
\providecommand \enquote  [1]{``#1''}%
\providecommand \bibnamefont  [1]{#1}%
\providecommand \bibfnamefont [1]{#1}%
\providecommand \citenamefont [1]{#1}%
\providecommand \href@noop [0]{\@secondoftwo}%
\providecommand \href [0]{\begingroup \@sanitize@url \@href}%
\providecommand \@href[1]{\@@startlink{#1}\@@href}%
\providecommand \@@href[1]{\endgroup#1\@@endlink}%
\providecommand \@sanitize@url [0]{\catcode `\\12\catcode `\$12\catcode
  `\&12\catcode `\#12\catcode `\^12\catcode `\_12\catcode `\%12\relax}%
\providecommand \@@startlink[1]{}%
\providecommand \@@endlink[0]{}%
\providecommand \url  [0]{\begingroup\@sanitize@url \@url }%
\providecommand \@url [1]{\endgroup\@href {#1}{\urlprefix }}%
\providecommand \urlprefix  [0]{URL }%
\providecommand \Eprint [0]{\href }%
\providecommand \doibase [0]{http://dx.doi.org/}%
\providecommand \selectlanguage [0]{\@gobble}%
\providecommand \bibinfo  [0]{\@secondoftwo}%
\providecommand \bibfield  [0]{\@secondoftwo}%
\providecommand \translation [1]{[#1]}%
\providecommand \BibitemOpen [0]{}%
\providecommand \bibitemStop [0]{}%
\providecommand \bibitemNoStop [0]{.\EOS\space}%
\providecommand \EOS [0]{\spacefactor3000\relax}%
\providecommand \BibitemShut  [1]{\csname bibitem#1\endcsname}%
\let\auto@bib@innerbib\@empty
\bibitem [{\citenamefont {Fayet}(1980{\natexlab{a}})}]{Fayet}%
  \BibitemOpen
  \bibfield  {author} {\bibinfo {author} {\bibfnamefont {P.}~\bibnamefont
  {Fayet}},\ }\href {\doibase 10.1016/0370-2693(80)90488-8} {\bibfield
  {journal} {\bibinfo  {journal} {Phys. Lett. B}\ }\textbf {\bibinfo {volume}
  {95}},\ \bibinfo {pages} {285} (\bibinfo {year}
  {1980}{\natexlab{a}})}\BibitemShut {NoStop}%
\bibitem [{\citenamefont {Fayet}(1980{\natexlab{b}})}]{Fayet2}%
  \BibitemOpen
  \bibfield  {author} {\bibinfo {author} {\bibfnamefont {P.}~\bibnamefont
  {Fayet}},\ }\href {\doibase 10.1016/0370-2693(80)90217-8} {\bibfield
  {journal} {\bibinfo  {journal} {Phys. Lett. B}\ }\textbf {\bibinfo {volume}
  {96}},\ \bibinfo {pages} {83} (\bibinfo {year}
  {1980}{\natexlab{b}})}\BibitemShut {NoStop}%
\bibitem [{\citenamefont {Arkani-Hamed}\ \emph {et~al.}(2009)\citenamefont
  {Arkani-Hamed}, \citenamefont {Finkbeiner}, \citenamefont {Slatyer},\ and\
  \citenamefont {Weiner}}]{lightV}%
  \BibitemOpen
  \bibfield  {author} {\bibinfo {author} {\bibfnamefont {N.}~\bibnamefont
  {Arkani-Hamed}}, \bibinfo {author} {\bibfnamefont {D.~P.}\ \bibnamefont
  {Finkbeiner}}, \bibinfo {author} {\bibfnamefont {T.~R.}\ \bibnamefont
  {Slatyer}}, \ and\ \bibinfo {author} {\bibfnamefont {N.}~\bibnamefont
  {Weiner}},\ }\href {\doibase 10.1103/PhysRevD.79.015014} {\bibfield
  {journal} {\bibinfo  {journal} {Phys. Rev. D}\ }\textbf {\bibinfo {volume}
  {79}},\ \bibinfo {pages} {015014} (\bibinfo {year} {2009})},\ \Eprint
  {http://arxiv.org/abs/0810.0713} {arXiv:0810.0713 [hep-ph]} \BibitemShut
  {NoStop}%
\bibitem [{\citenamefont {Pospelov}\ and\ \citenamefont
  {Ritz}(2009)}]{lightV2}%
  \BibitemOpen
  \bibfield  {author} {\bibinfo {author} {\bibfnamefont {M.}~\bibnamefont
  {Pospelov}}\ and\ \bibinfo {author} {\bibfnamefont {A.}~\bibnamefont
  {Ritz}},\ }\href {\doibase 10.1016/j.physletb.2008.12.012} {\bibfield
  {journal} {\bibinfo  {journal} {Phys. Lett. B}\ }\textbf {\bibinfo {volume}
  {671}},\ \bibinfo {pages} {391} (\bibinfo {year} {2009})},\ \Eprint
  {http://arxiv.org/abs/0810.1502} {arXiv:0810.1502 [hep-ph]} \BibitemShut
  {NoStop}%
\bibitem [{\citenamefont {Merkel}\ \emph {et~al.}(2011)\citenamefont {Merkel}
  \emph {et~al.}}]{A1}%
  \BibitemOpen
  \bibfield  {author} {\bibinfo {author} {\bibfnamefont {H.}~\bibnamefont
  {Merkel}} \emph {et~al.} (\bibinfo {collaboration} {A1 Collaboration}),\
  }\href {\doibase 10.1103/PhysRevLett.106.251802} {\bibfield  {journal}
  {\bibinfo  {journal} {Phys. Rev. Lett.}\ }\textbf {\bibinfo {volume} {106}},\
  \bibinfo {pages} {251802} (\bibinfo {year} {2011})},\ \Eprint
  {http://arxiv.org/abs/1101.4091} {arXiv:1101.4091 [nucl-ex]} \BibitemShut
  {NoStop}%
\bibitem [{\citenamefont {Abrahamyan}\ \emph {et~al.}(2011)\citenamefont
  {Abrahamyan} \emph {et~al.}}]{Apex}%
  \BibitemOpen
  \bibfield  {author} {\bibinfo {author} {\bibfnamefont {S.}~\bibnamefont
  {Abrahamyan}} \emph {et~al.} (\bibinfo {collaboration} {APEX
  Collaboration}),\ }\href {\doibase 10.1103/PhysRevLett.107.191804} {\bibfield
   {journal} {\bibinfo  {journal} {Phys. Rev. Lett.}\ }\textbf {\bibinfo
  {volume} {107}},\ \bibinfo {pages} {191804} (\bibinfo {year} {2011})},\
  \Eprint {http://arxiv.org/abs/1108.2750} {arXiv:1108.2750 [hep-ex]}
  \BibitemShut {NoStop}%
\bibitem [{\citenamefont {Bennett}\ \emph {et~al.}(2006)\citenamefont {Bennett}
  \emph {et~al.}}]{g-2}%
  \BibitemOpen
  \bibfield  {author} {\bibinfo {author} {\bibfnamefont {G.}~\bibnamefont
  {Bennett}} \emph {et~al.} (\bibinfo {collaboration} {Muon G-2
  Collaboration}),\ }\href {\doibase 10.1103/PhysRevD.73.072003} {\bibfield
  {journal} {\bibinfo  {journal} {Phys. Rev. D}\ }\textbf {\bibinfo {volume}
  {73}},\ \bibinfo {pages} {072003} (\bibinfo {year} {2006})},\ \Eprint
  {http://arxiv.org/abs/hep-ex/0602035} {arXiv:hep-ex/0602035 [hep-ex]}
  \BibitemShut {NoStop}%
\bibitem [{\citenamefont {Pohl}\ \emph {et~al.}(2010)\citenamefont {Pohl},
  \citenamefont {Antognini}, \citenamefont {Nez}, \citenamefont {Amaro},
  \citenamefont {Biraben} \emph {et~al.}}]{muH}%
  \BibitemOpen
  \bibfield  {author} {\bibinfo {author} {\bibfnamefont {R.}~\bibnamefont
  {Pohl}}, \bibinfo {author} {\bibfnamefont {A.}~\bibnamefont {Antognini}},
  \bibinfo {author} {\bibfnamefont {F.}~\bibnamefont {Nez}}, \bibinfo {author}
  {\bibfnamefont {F.~D.}\ \bibnamefont {Amaro}}, \bibinfo {author}
  {\bibfnamefont {F.}~\bibnamefont {Biraben}},  \emph {et~al.},\ }\href
  {\doibase 10.1038/nature09250} {\bibfield  {journal} {\bibinfo  {journal}
  {Nature}\ }\textbf {\bibinfo {volume} {466}},\ \bibinfo {pages} {213}
  (\bibinfo {year} {2010})}\BibitemShut {NoStop}%
\bibitem [{\citenamefont {Mohr}\ \emph {et~al.}(2008)\citenamefont {Mohr},
  \citenamefont {Taylor},\ and\ \citenamefont {Newell}}]{CODATA}%
  \BibitemOpen
  \bibfield  {author} {\bibinfo {author} {\bibfnamefont {P.~J.}\ \bibnamefont
  {Mohr}}, \bibinfo {author} {\bibfnamefont {B.~N.}\ \bibnamefont {Taylor}}, \
  and\ \bibinfo {author} {\bibfnamefont {D.~B.}\ \bibnamefont {Newell}},\
  }\href {\doibase 10.1103//RevModPhys.80.633} {\bibfield  {journal} {\bibinfo
  {journal} {Rev. Mod. Phys.}\ }\textbf {\bibinfo {volume} {80}},\ \bibinfo
  {pages} {633} (\bibinfo {year} {2008})},\ \Eprint
  {http://arxiv.org/abs/0801.0028} {arXiv:0801.0028 [physics.atom-ph]}
  \BibitemShut {NoStop}%
\bibitem [{\citenamefont {Hill}\ and\ \citenamefont {Paz}(2010)}]{HillPaz}%
  \BibitemOpen
  \bibfield  {author} {\bibinfo {author} {\bibfnamefont {R.~J.}\ \bibnamefont
  {Hill}}\ and\ \bibinfo {author} {\bibfnamefont {G.}~\bibnamefont {Paz}},\
  }\href {\doibase 10.1103/PhysRevD.82.113005} {\bibfield  {journal} {\bibinfo
  {journal} {Phys. Rev. D}\ }\textbf {\bibinfo {volume} {82}},\ \bibinfo
  {pages} {113005} (\bibinfo {year} {2010})},\ \Eprint
  {http://arxiv.org/abs/1008.4619} {arXiv:1008.4619 [hep-ph]} \BibitemShut
  {NoStop}%
\bibitem [{\citenamefont {Okun}(1982)}]{Okun}%
  \BibitemOpen
  \bibfield  {author} {\bibinfo {author} {\bibfnamefont {L.}~\bibnamefont
  {Okun}},\ }\href@noop {} {\bibfield  {journal} {\bibinfo  {journal} {Sov.
  Phys. JETP}\ }\textbf {\bibinfo {volume} {56}},\ \bibinfo {pages} {502}
  (\bibinfo {year} {1982})}\BibitemShut {NoStop}%
\bibitem [{\citenamefont {Holdom}(1986)}]{Holdom}%
  \BibitemOpen
  \bibfield  {author} {\bibinfo {author} {\bibfnamefont {B.}~\bibnamefont
  {Holdom}},\ }\href {\doibase 10.1016/0370-2693(86)91377-8} {\bibfield
  {journal} {\bibinfo  {journal} {Phys. Lett. B}\ }\textbf {\bibinfo {volume}
  {166}},\ \bibinfo {pages} {196} (\bibinfo {year} {1986})}\BibitemShut
  {NoStop}%
\bibitem [{\citenamefont {Gninenko}\ and\ \citenamefont
  {Krasnikov}(2001)}]{Gninenko}%
  \BibitemOpen
  \bibfield  {author} {\bibinfo {author} {\bibfnamefont {S.}~\bibnamefont
  {Gninenko}}\ and\ \bibinfo {author} {\bibfnamefont {N.}~\bibnamefont
  {Krasnikov}},\ }\href {\doibase 10.1016/S0370-2693(01)00693-1} {\bibfield
  {journal} {\bibinfo  {journal} {Phys. Lett. B}\ }\textbf {\bibinfo {volume}
  {513}},\ \bibinfo {pages} {119} (\bibinfo {year} {2001})},\ \Eprint
  {http://arxiv.org/abs/hep-ph/0102222} {arXiv:hep-ph/0102222 [hep-ph]}
  \BibitemShut {NoStop}%
\bibitem [{\citenamefont {Pospelov}(2009)}]{Pospelov}%
  \BibitemOpen
  \bibfield  {author} {\bibinfo {author} {\bibfnamefont {M.}~\bibnamefont
  {Pospelov}},\ }\href {\doibase 10.1103/PhysRevD.80.095002} {\bibfield
  {journal} {\bibinfo  {journal} {Phys. Rev. D}\ }\textbf {\bibinfo {volume}
  {80}},\ \bibinfo {pages} {095002} (\bibinfo {year} {2009})},\ \Eprint
  {http://arxiv.org/abs/0811.1030} {arXiv:0811.1030 [hep-ph]} \BibitemShut
  {NoStop}%
\bibitem [{\citenamefont {Jaeckel}\ and\ \citenamefont
  {Roy}(2010)}]{newforce1}%
  \BibitemOpen
  \bibfield  {author} {\bibinfo {author} {\bibfnamefont {J.}~\bibnamefont
  {Jaeckel}}\ and\ \bibinfo {author} {\bibfnamefont {S.}~\bibnamefont {Roy}},\
  }\href {\doibase 10.1103/PhysRevD.82.125020} {\bibfield  {journal} {\bibinfo
  {journal} {Phys. Rev. D}\ }\textbf {\bibinfo {volume} {82}},\ \bibinfo
  {pages} {125020} (\bibinfo {year} {2010})},\ \Eprint
  {http://arxiv.org/abs/1008.3536} {arXiv:1008.3536 [hep-ph]} \BibitemShut
  {NoStop}%
\bibitem [{\citenamefont {Barger}\ \emph {et~al.}(2011)\citenamefont {Barger},
  \citenamefont {Chiang}, \citenamefont {Keung},\ and\ \citenamefont
  {Marfatia}}]{newforce2}%
  \BibitemOpen
  \bibfield  {author} {\bibinfo {author} {\bibfnamefont {V.}~\bibnamefont
  {Barger}}, \bibinfo {author} {\bibfnamefont {C.-W.}\ \bibnamefont {Chiang}},
  \bibinfo {author} {\bibfnamefont {W.-Y.}\ \bibnamefont {Keung}}, \ and\
  \bibinfo {author} {\bibfnamefont {D.}~\bibnamefont {Marfatia}},\ }\href
  {\doibase 10.1103/PhysRevLett.106.153001} {\bibfield  {journal} {\bibinfo
  {journal} {Phys. Rev. Lett.}\ }\textbf {\bibinfo {volume} {106}},\ \bibinfo
  {pages} {153001} (\bibinfo {year} {2011})},\ \Eprint
  {http://arxiv.org/abs/1011.3519} {arXiv:1011.3519 [hep-ph]} \BibitemShut
  {NoStop}%
\bibitem [{\citenamefont {Tucker-Smith}\ and\ \citenamefont
  {Yavin}(2011)}]{newforce3}%
  \BibitemOpen
  \bibfield  {author} {\bibinfo {author} {\bibfnamefont {D.}~\bibnamefont
  {Tucker-Smith}}\ and\ \bibinfo {author} {\bibfnamefont {I.}~\bibnamefont
  {Yavin}},\ }\href {\doibase 10.1103/PhysRevD.83.101702} {\bibfield  {journal}
  {\bibinfo  {journal} {Phys. Rev. D}\ }\textbf {\bibinfo {volume} {83}},\
  \bibinfo {pages} {101702} (\bibinfo {year} {2011})},\ \Eprint
  {http://arxiv.org/abs/1011.4922} {arXiv:1011.4922 [hep-ph]} \BibitemShut
  {NoStop}%
\bibitem [{\citenamefont {Batell}\ \emph {et~al.}(2011)\citenamefont {Batell},
  \citenamefont {McKeen},\ and\ \citenamefont {Pospelov}}]{BMP}%
  \BibitemOpen
  \bibfield  {author} {\bibinfo {author} {\bibfnamefont {B.}~\bibnamefont
  {Batell}}, \bibinfo {author} {\bibfnamefont {D.}~\bibnamefont {McKeen}}, \
  and\ \bibinfo {author} {\bibfnamefont {M.}~\bibnamefont {Pospelov}},\ }\href
  {\doibase 10.1103/PhysRevLett.107.011803} {\bibfield  {journal} {\bibinfo
  {journal} {Phys. Rev. Lett.}\ }\textbf {\bibinfo {volume} {107}},\ \bibinfo
  {pages} {011803} (\bibinfo {year} {2011})},\ \Eprint
  {http://arxiv.org/abs/1103.0721} {arXiv:1103.0721 [hep-ph]} \BibitemShut
  {NoStop}%
\bibitem [{\citenamefont {Missimer}\ and\ \citenamefont {Simons}(1985)}]{MS}%
  \BibitemOpen
  \bibfield  {author} {\bibinfo {author} {\bibfnamefont {J.~H.}\ \bibnamefont
  {Missimer}}\ and\ \bibinfo {author} {\bibfnamefont {L.~M.}\ \bibnamefont
  {Simons}},\ }\href {\doibase 10.1016/0370-1573(85)90013-4} {\bibfield
  {journal} {\bibinfo  {journal} {Phys. Rept.}\ }\textbf {\bibinfo {volume}
  {118}},\ \bibinfo {pages} {179} (\bibinfo {year} {1985})}\BibitemShut
  {NoStop}%
\bibitem [{\citenamefont {Feinberg}\ and\ \citenamefont {Chen}(1974)}]{CF}%
  \BibitemOpen
  \bibfield  {author} {\bibinfo {author} {\bibfnamefont {G.}~\bibnamefont
  {Feinberg}}\ and\ \bibinfo {author} {\bibfnamefont {M.}~\bibnamefont
  {Chen}},\ }\href {\doibase 10.1103/PhysRevD.10.190, 10.1103/PhysRevD.10.3145}
  {\bibfield  {journal} {\bibinfo  {journal} {Phys. Rev. D}\ }\textbf {\bibinfo
  {volume} {10}},\ \bibinfo {pages} {190} (\bibinfo {year} {1974})},\ \bibinfo
  {note} {[Erratum-ibid.\ D {\bf 10}, 3145 (1974)]}\BibitemShut {NoStop}%
\bibitem [{\citenamefont {Kirch}\ \emph {et~al.}(1997)\citenamefont {Kirch}
  \emph {et~al.}}]{Klaus}%
  \BibitemOpen
  \bibfield  {author} {\bibinfo {author} {\bibfnamefont {K.}~\bibnamefont
  {Kirch}} \emph {et~al.},\ }\href {\doibase 10.1103/PhysRevLett.78.4363}
  {\bibfield  {journal} {\bibinfo  {journal} {Phys. Rev. Lett.}\ }\textbf
  {\bibinfo {volume} {78}},\ \bibinfo {pages} {4363} (\bibinfo {year}
  {1997})}\BibitemShut {NoStop}%
\bibitem [{\citenamefont {Grechukhin}\ and\ \citenamefont
  {Soldatov}(1979)}]{GS}%
  \BibitemOpen
  \bibfield  {author} {\bibinfo {author} {\bibfnamefont {D.}~\bibnamefont
  {Grechukhin}}\ and\ \bibinfo {author} {\bibfnamefont {A.}~\bibnamefont
  {Soldatov}},\ }\href@noop {} {\bibfield  {journal} {\bibinfo  {journal} {Sov.
  Phys. JETP}\ }\textbf {\bibinfo {volume} {50}},\ \bibinfo {pages} {1039}
  (\bibinfo {year} {1979})}\BibitemShut {NoStop}%
\bibitem [{\citenamefont {Hartmann}\ \emph {et~al.}(1982)\citenamefont
  {Hartmann} \emph {et~al.}}]{cascade1}%
  \BibitemOpen
  \bibfield  {author} {\bibinfo {author} {\bibfnamefont {F.~J.}\ \bibnamefont
  {Hartmann}} \emph {et~al.},\ }\href {http://dx.doi.org/10.1007/BF01417434}
  {\bibfield  {journal} {\bibinfo  {journal} {Z. Phys. A}\ }\textbf {\bibinfo
  {volume} {305}},\ \bibinfo {pages} {189} (\bibinfo {year}
  {1982})}\BibitemShut {NoStop}%
\bibitem [{\citenamefont {Naumann}\ \emph {et~al.}(1985)\citenamefont {Naumann}
  \emph {et~al.}}]{cascade2}%
  \BibitemOpen
  \bibfield  {author} {\bibinfo {author} {\bibfnamefont {R.~A.}\ \bibnamefont
  {Naumann}} \emph {et~al.},\ }\href {\doibase 10.1103/PhysRevA.31.727}
  {\bibfield  {journal} {\bibinfo  {journal} {Phys. Rev. A}\ }\textbf {\bibinfo
  {volume} {31}},\ \bibinfo {pages} {727} (\bibinfo {year} {1985})}\BibitemShut
  {NoStop}%
\bibitem [{\citenamefont {Bethe}\ and\ \citenamefont {Salpeter}(1957)}]{BS}%
  \BibitemOpen
  \bibfield  {author} {\bibinfo {author} {\bibfnamefont {H.~A.}\ \bibnamefont
  {Bethe}}\ and\ \bibinfo {author} {\bibfnamefont {E.~E.}\ \bibnamefont
  {Salpeter}},\ }\href@noop {} {\emph {\bibinfo {title} {Quantum Mechanics of
  One- and Two-Electron Atoms}}}\ (\bibinfo  {publisher} {Springer-Verlag, New
  York},\ \bibinfo {year} {1957})\BibitemShut {NoStop}%
\bibitem [{\citenamefont {Berestetskii}\ \emph {et~al.}(1982)\citenamefont
  {Berestetskii}, \citenamefont {Lifshitz},\ and\ \citenamefont
  {Pitaevskii}}]{LL}%
  \BibitemOpen
  \bibfield  {author} {\bibinfo {author} {\bibfnamefont {V.}~\bibnamefont
  {Berestetskii}}, \bibinfo {author} {\bibfnamefont {E.}~\bibnamefont
  {Lifshitz}}, \ and\ \bibinfo {author} {\bibfnamefont {L.}~\bibnamefont
  {Pitaevskii}},\ }\href@noop {} {\emph {\bibinfo {title} {Quantum
  Electrodynamics}}},\ \bibinfo {edition} {2nd}\ ed.\ (\bibinfo  {publisher}
  {Pergamon, Oxford},\ \bibinfo {year} {1982})\BibitemShut {NoStop}%
\bibitem [{\citenamefont {Chatterjee}\ \emph {et~al.}(1992)\citenamefont
  {Chatterjee}, \citenamefont {Das},\ and\ \citenamefont
  {Chakraborty}}]{Chatt}%
  \BibitemOpen
  \bibfield  {author} {\bibinfo {author} {\bibfnamefont {L.}~\bibnamefont
  {Chatterjee}}, \bibinfo {author} {\bibfnamefont {G.}~\bibnamefont {Das}}, \
  and\ \bibinfo {author} {\bibfnamefont {A.}~\bibnamefont {Chakraborty}},\
  }\href {\doibase 10.1209/0295-5075/18/2/010} {\bibfield  {journal} {\bibinfo
  {journal} {Europhys. Lett.}\ }\textbf {\bibinfo {volume} {18}},\ \bibinfo
  {pages} {145} (\bibinfo {year} {1992})}\BibitemShut {NoStop}%
\end{thebibliography}%

\end{document}